\documentclass{aa}
\usepackage{psfig}
\usepackage{graphicx}

\def\bron{2S~0918--549}
\def\ecs{erg~cm$^{-2}$s$^{-1}$}
\def\lum{erg~s$^{-1}$}
\newcommand{\gtap}{\mathrel{\hbox{\rlap{\lower.55ex \hbox {$\sim$}}
                    \kern-.3em \raise.4ex \hbox{$>$}}}}
\newcommand{\ltap}{\mathrel{\hbox{\rlap{\lower.55ex \hbox {$\sim$}}
                    \kern-.3em \raise.4ex \hbox{$<$}}}}

\begin{document}

\title{On the possibility of a helium white dwarf donor in
the presumed ultracompact binary \bron}

\titlerunning{A helium white dwarf donor in \bron}
\authorrunning{J.J.M. in 't Zand, A. Cumming, M. van der Sluys et al.}

\author{
J.J.M.~in~'t~Zand\inst{1,2},
A. Cumming\inst{3},
M.V. van der Sluys\inst{2},
F.~Verbunt\inst{2} \&
O.R. Pols\inst{2}
}


\institute{     SRON National Institute for Space Research, Sorbonnelaan 2,
                NL - 3584 CA Utrecht, the Netherlands 
	 \and
                Astronomical Institute, Utrecht University, P.O. Box 80000,
                NL - 3508 TA Utrecht, the Netherlands
         \and
                Physics Department, McGill University, 3600 Rue University,
                Montreal, QC, H3A 2T8, Canada
	}

\date{Accepted June 19, 2005}

\abstract{\bron\ is a low-mass X-ray binary (LMXB) with a low optical
 to X-ray flux ratio. Probably it is an ultracompact binary with an
 orbital period shorter than 60~min. Such binaries cannot harbor
 hydrogen rich donor stars. As with other (sometimes confirmed)
 ultracompact LMXBs, \bron\ is observed to have a high
 neon-to-oxygen abundance ratio (Juett et al. 2001) which has been
 used to argue that the companion star is a CO or ONe white
 dwarf. However, type-I X-ray bursts have been observed from several
 of these systems implying the presence of hydrogen or helium on the
 neutron star surface. In this paper, we argue that the companion star
 in \bron\ is a helium white dwarf We first present a Type I X-ray
 burst from \bron\ with a long duration of 40 minutes. We show
 that this burst is naturally explained by accretion of pure helium at
 the inferred accretion rate of $\sim 0.01$ times the Eddington
 accretion rate. At higher accretion rates of $\sim 0.1$ Eddington,
 hydrogen is required to explain long duration bursts. However, at low
 rates the long duration is due to the large amount of helium that
 accumulates prior to the burst. We show that it is possible to form a
 helium white dwarf donor in an ultracompact binary if accretion
 starts during the first ascent of the giant branch, when the core is
 made of predominantly helium. Furthermore, this scenario naturally
 explains the high neon-to-oxgen ratio, without requiring a CO or ONe
 white dwarf companion.  The only observational aspect of \bron\ that
 we cannot explain is the absence of helium lines in the optical
 spectrum. Model calculations of optical accretion disk spectra need
 to be carried out in order to obtain limits on the helium abundance.
 }

\maketitle 

\section{Introduction}
\label{intro}

\bron\ is a low-mass X-ray binary (LMXB) in which a Roche-lobe filling
low-mass star orbits a neutron star and the liberation of
gravitational energy of the in-falling matter produces X-rays.  It is a
particularly interesting example of a LMXB for three reasons. First,
it has an optical to X-ray flux ratio which suggests that it is an
ultracompact binary with an orbital period less than $\sim$60~min and
a companion donor star that can only fit within the Roche lobe if it
is hydrogen depleted (Juett et al. 2001, based on relations
established by Van Paradijs \& McClintock 1994). This was recently
confirmed through optical spectroscopy which revealed the lack of
lines from hydrogen and helium (Nelemans et al. 2004). It suggests
that the companion star is a C-O or O-Ne white dwarf (cf., Nelson,
Rappaport \& Joss 1986).

Second, it has an X-ray luminosity which is low for an active LMXB,
namely less than or equal to 1\% of the Eddington limit for a
canonical (1.4~$M_\odot$, 10 km radius) neutron star (Jonker et
al. 2001).

Third, \bron\ appears to have an unusually high Ne/O abundance ratio,
a characteristic which it shares with three other non-pulsating LMXBs
out of 56 cases investigated by Juett et al. (2001). Juett et
al. propose that this is related to a degenerate donor in the
suggested ultracompact nature of the binary. This proposition is
supported by detections of orbital periods in two of the other three
(18 min in 4U 1543-624, see Wang \& Chakrabarty 2004; 21 min in 4U
1850-087, see Homer et al. 1996), although those detections need
corroboration since they were made only once in each case. Recently,
it was observed for two of these high Ne/O systems that the Ne/O ratio
changed while for another previously not measured ultracompact system
the Ne/O ratio was not high (Juett \& Chakrabarty 2003 and 2005). This
indicates that the interpretation of Ne/O in terms of donor
evolutionary status is less straightforward and is possibly biased by
ionization effects. For \bron, nevertheless, the Ne/O ratio was
found be consistently 2.4 times the solar value, as measured with
ASCA, Chandra-LETGS and XMM-Newton observations, while the flux
changed almost by an order of magnitude.

Three of the four LMXBs with a possibly high Ne/O-ratio exhibit
sporadic type-I X-ray bursts which are due to thermonuclear flashes in
the upper layers of a neutron star. Two have been reported from
4U~0614+09 (Swank et al. 1978; Brandt et al. 1992), four from
4U~1850-087 (Swank et al. 1976; Cominsky et al. 1977 and 1981) and
three from \bron\ (Jonker et al. 2001; Cornelisse et al. 2002;
Galloway et al. 2005). We here report four additional bursts from the
latter. Additional bursts for the other systems were also detected,
particularly with HETE-II (see for instance Nakagawa et al. 2004), but
these are not published in detail yet. The one source never seen
bursting is 4U~1543-624.

As noted by Juett et al. (2003) and Nelemans et al. (2004), there is a
puzzling contradiction between the characteristics (or mere presence)
of these bursts and the suggested hydrogen or helium depletion
in the donor stars. The bursts detected from these systems last
between 10 and a few hundred seconds, suggesting a high hydrogen and
helium content in the flash fuel. The question is: where does the
hydrogen and helium come from if not from the donor star? The problem
is more severe for hydrogen than for helium because in an evolved
donor star like in an ultracompact system the hydrogen can reside only
in the outer layers that were lost being outside the Roche lobe. Juett
et al. and Nelemans et al. suggest that spallation of accreted
elements may be important (e.g., Bildsten, Salpeter \& Wasserman
1992). However, it is not trivial to invoke spallation. Spallation
requires hydrogen nuclei (protons) to bombard the higher-up
Coulomb-stopped heavy nuclei and create lighter elements, but the
problem is that protons are in short supply. A definite assessment of
the viability of this process needs to come from new calculations that
also take into account non-radial accretion (e.g., Bildsten, Chang \&
Paerels 2003) and high metal abundances.

In this paper we present (in sections \ref{synopsis} and
\ref{analong}) measurements of an extraordinary burst from \bron\
which was detected with the Wide Field Cameras (WFCs; Jager et
al. 1997) on board BeppoSAX (Boella et al. 1997a). It is the longest
of all bursts observed from any (presumed) ultracompact, lasting over
half an hour. Commonly, long burst durations (for example,
bursts from the regular burster GS~1826-24; Ubertini et al. 1999;
Galloway et al. 2004) are explained by a high hydrogen fuel
content. The protons are captured by the ashes of unstable helium
burning and initiate a relatively slow beta decay process (the rp
process) that is responsible for the burst longevity (e.g., Fujimoto
et al. 1981). However, at the low accretion rate of $\sim 1$\%
Eddington appropriate for \bron, the conditions at the time of
ignition of the flash are different. Because the fuel accumulates
slowly, any hydrogen has time to stably burn away, leaving a thick
layer of helium which ignites and burns in a long duration and
energetic burst. In fact, given the likely ultracompact nature of this
source, we argue that the long duration burst is due to accretion of
pure helium from a helium white dwarf companion. As we show in section
\ref{dis1}, \bron\ provides the rarely seen circumstances for a
long duration helium flash to be possible, being a persistent X-ray
source at a fairly low mass accretion rate. We investigate in
section \ref{dis2} evolutionary paths to arrive at the implied
helium-rich donor star and find a likely path leading to a helium
white dwarf. What is more, this star is predicted to have a Ne/O
overabundance ratio which confirms the observations. Thus, a
model in which the companion star is a helium white dwarf explains
many peculiar details about \bron. One detail which is not explained
concerns the lack of helium lines in the optical spectrum.

\begin{table}[t]
\caption[]{List of X-ray bursts from \bron \label{tab1}. Bursts 2, 3
and 7 have been published before. Values between parentheses represent
uncertainties in the last digit.}
\begin{tabular}{llllll}
\hline\hline 
No & Date & Instr.     & Peak         & Persistent & $\tau$ \\
   &  &                & flux         & flux       &        \\
   &  &                & \multicolumn{2}{c}{(Crab units$^\star$)}  &(sec)   \\
\hline
1         & 1996 Oct 1  & WFC  &  3.7(3) & 0.0048(5) & 117(2)     \\
2$^\dag$  & 1999 Jun 10 & WFC  &  3.3(3) & 0.0037(5) & 29(4)      \\
3$^\ddag$ & 2000 May 12 & PCA  &  $>3.2$ & $<0.0146$ & 8.95(5) \\
4         & 2001 May 18 & WFC  &  2.5(3) & 0.0094(5) & 5.8(4)   \\
5         & 2001 Sep 29 & ASM  &  2.9(1) & 0.0061(7)$^\P$ & 25(3)   \\
6         & 2003 Aug 5  & ASM  &  2.1(1) & 0.0080(8)$^\P$ & 22(10)$^\times$   \\
7$^\ast$ & 2004 Jun 18& PCA& 0.26(1)& 0.0126(3) & 12.5(5) \\
\hline\hline
\multicolumn{6}{l}{$^\star$\parbox[t]{0.9\columnwidth}{For a burst peak spectrum as
determined in this paper for the first burst, one Crab unit translates to a bolometric
flux of $2.7\times10^{-8}$~\ecs. For the persistent flux, one may adhere to a 2--10 keV
flux of $2.0\times10^{-8}$~\ecs\ (the bolometric correction is less certain in that case).}}\\
\multicolumn{6}{l}{$^\dag$\parbox[t]{0.9\columnwidth}{Cornelisse et al. 2002. This burst was
erroneously dated four days earlier in that paper.}}\\
\multicolumn{6}{l}{$^\ddag$Jonker et al. 2001}\\
\multicolumn{6}{l}{$^\ast$\parbox[t]{0.9\columnwidth}{Galloway et al. 2005 (this paper mentioned 2 more faint
bursts from \bron, but the data do not allow
confirmation as type-I X-ray bursts}}\\
\multicolumn{6}{l}{$^\P$These are 14-d averages.}\\
\multicolumn{6}{l}{$^\times$\parbox[t]{0.9\columnwidth}{This burst was only partly observed. The observation stopped 18 s after burst onset.}}\\
\end{tabular}
\end{table}

\begin{figure}[t]
\includegraphics[width=\columnwidth,angle=0]{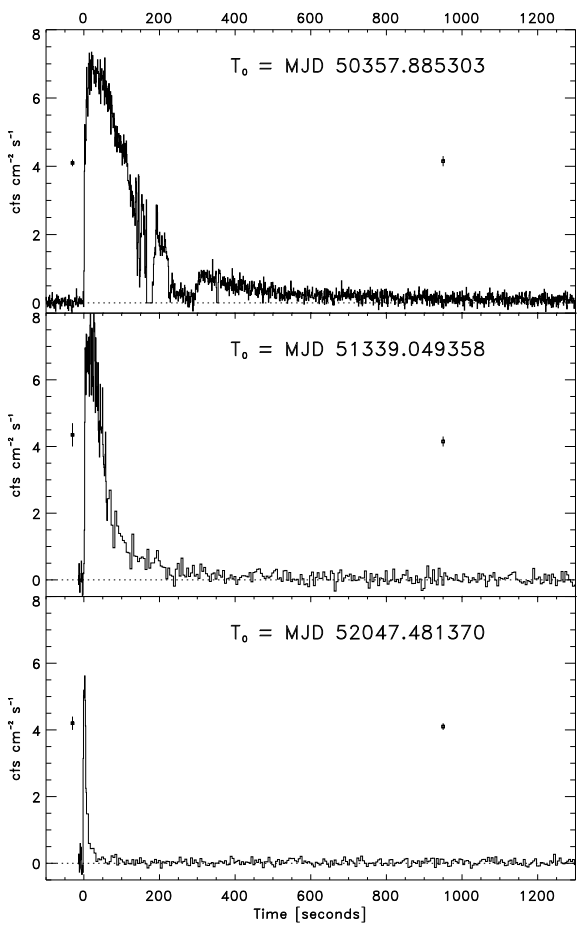}
\caption[]{2-28 keV light curves of the three WFC-detected bursts at
varying time resolution. Typical error bars are indicated at the left and
right of each panel.
\label{figthree}}
\end{figure}

\section{Synopsis of X-ray bursts from \bron}
\label{synopsis}

Thus far three X-ray bursts were reported from \bron, by Jonker et
al. (2001), Cornelisse et al. (2002) and Galloway et al. (2005). The
first two bursts have similar bolometric peak fluxes of 8.8 and
9.4$\times10^{-8}$~\ecs, but the decay times differ by a factor of 3
(see Table~\ref{tab1}). The third burst as identified by Galloway is
an order of magnitude fainter. We carried out archival searches in
BeppoSAX/WFC data (net exposure 9.2 Msec), RXTE/ASM (2.6 Msec for an
effective exposure time of 70~s per dwell), and RXTE/PCA data
($\sim$300 ksec, including data from AO9), and found 2 more bursts in
ASM and 2 in WFC data. No reports of bursts were made from
observations with Einstein, EXOSAT, ROSAT, ASCA, BeppoSAX, Chandra and
XMM-Newton within a total of approximately 170 ksec. We derive an
average burst rate (from simple division of the exposure time by
the number of bursts) of once every 20 days, but note that the WFCs
and ASM are not sensitive enough to detect the fainter bursts. Based
on PCA data alone, the average burst rate is once every $2\pm1$~d.
Table~\ref{tab1} reports the main characteristics of all 7 bursts.

\begin{figure}[t]
\includegraphics[height=\columnwidth,angle=90]{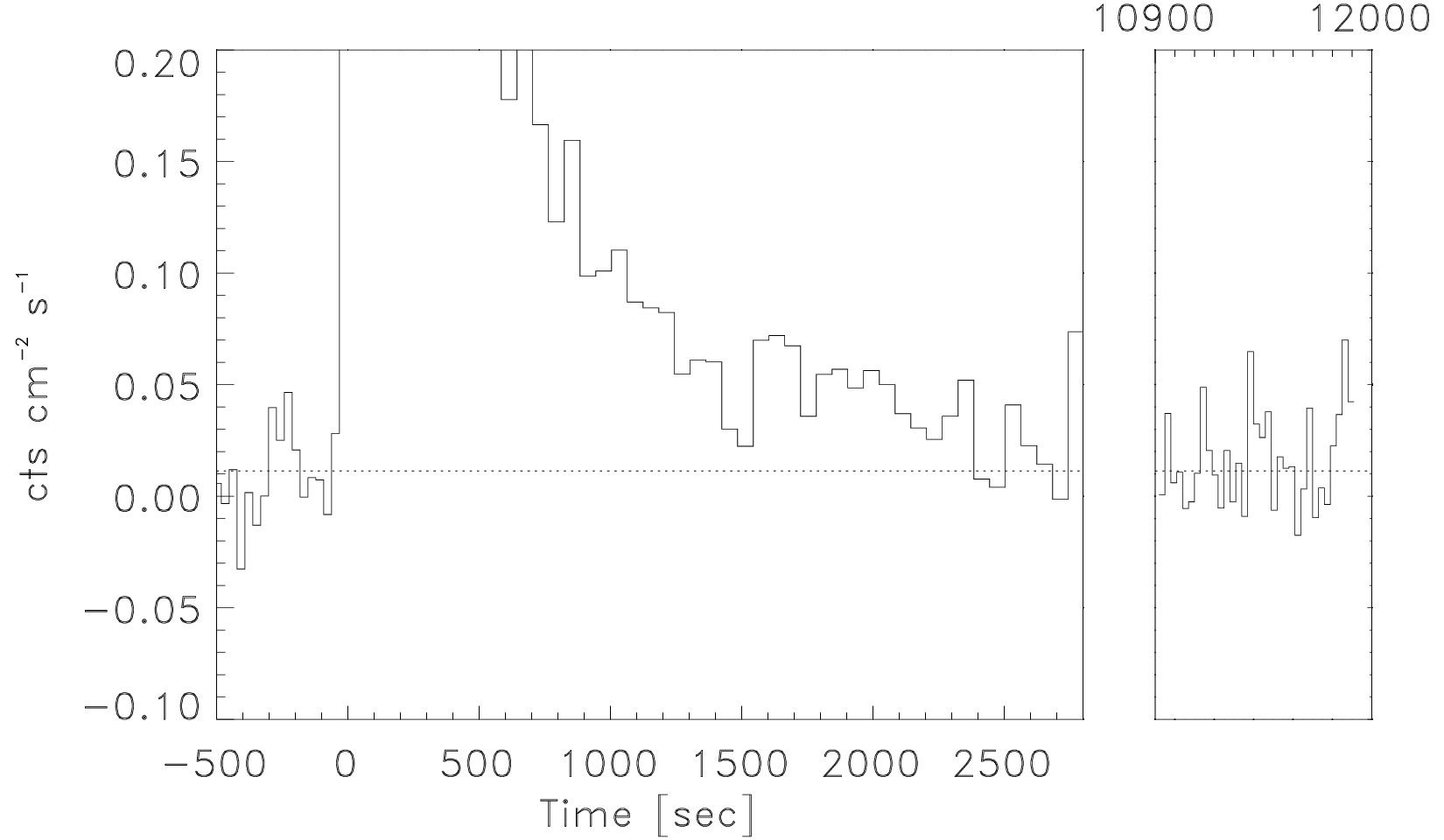}
\caption[]{2-28 keV light curve of the long burst, zooming in at low
flux levels and with a time resolution of 30 s. The dotted line
indicates the out-of-burst persistent flux level. There are no data
during times between the two panels.
\label{figone}}
\end{figure}

Except for the last, all bursts are fairly bright with bolometric peak
fluxes that translate to between $6\times10^{-8}$ and
10$^{-7}$~\ecs. Cornelisse et al. (2002), equalizing the peak flux of
the second burst to the Eddington limit of a hydrogen-rich
photosphere, derive a distance of 4.2~kpc with an uncertainty of
30\%. This is in contrast to an earlier distance estimate, from
optical measurements, of 15 kpc by Chevalier \& Ilovaisky (1987) which
is derived assuming the optical counterpart to be as luminous as in
other LMXBs. Therefore, the counterpart must be considerably sub
luminous which is indicative of a small accretion disk and short
orbital period of at most 60~min (Juett et al. 2001).

In Fig.~\ref{figthree} the three bursts detected with the WFC are
plotted on identical scales. This illustrates the diversity of the
bursts.  In particular it illustrates the longevity of the first
burst.  In Fig.~\ref{figone} the flux scale is blown up for this burst
and it is clear that it persists for at least approximately
2500~s. The second burst lasts 10 times as short, the third burst
nearly 100 times as short, the bright PCA-detected burst about 25
times (Jonker et al. 2001).

\section{Analysis of the long burst}
\label{analong}

The long burst started on October 1, 1996, at 21:14:51 UT, rose to
peak levels within 1~s and carried on for approximately 40~min before
it disappeared in the background noise (at a level $\approx$350 times
below the burst peak; see Fig.~\ref{figone}). The e-folding decay time
over the first 200 s is $117\pm2$~s. This is the longest decay time of
all 2427 type-I X-ray bursts measured with the WFCs that are not
superbursts (In 't Zand et al. 2004b; Kuulkers 2004), except
for one burst from SLX~1737-282 (In~'t~Zand et al. 2002) which
exhibited a decay time of 600~s.

The off-axis angle of \bron\ in the WFC field of view during the long
burst was near to optimum.  The source illuminated 90\% of the
available sensitive detector area. However, the observation was
plagued by high telemetry rates due to a bright source within the
concurrent narrow-field instruments, which resulted in sporadic WFC
data drop outs. Two drop outs occurred during the long X-ray burst,
from 165 to 180 sec after burst onset and from 350 to 354 sec.

\begin{figure}[!t]
\includegraphics[height=\columnwidth,angle=90]{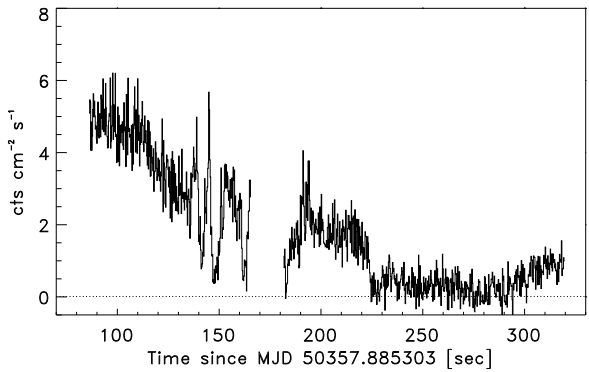}
\caption[]{2-28 keV light curve of the long burst, zooming in on the
period of strong variability at a time resolution of 0.25~s.
\label{figburstvar}}
\end{figure}

There are two more conspicuous features in the time profile of the
long burst, see Fig.~\ref{figburstvar}: two minutes after the burst
onset a 30~s period of strong variability is observed, with dips and
peaks that grow above the flux before that. Two minutes later the flux
suddenly (within 2 s) decreases by a factor of about
four. Subsequently it remains on a decaying track. After a little over
70~s it rises back to the extrapolated pre-drop downward trend and
continues its decay. Since several data drop outs occurred during this
observation we checked whether this drop in flux could be due to
telemetry overflow. We studied the flux history of another bright
source in the field of view, Vela X-1, and that of the remaining
background. These showed no flux decrease whatsoever, in contrast to
during the data drop out periods. We conclude that the flux decrease
during the burst is genuinely associated with \bron.

\begin{figure}[t]
\includegraphics[width=\columnwidth,angle=0]{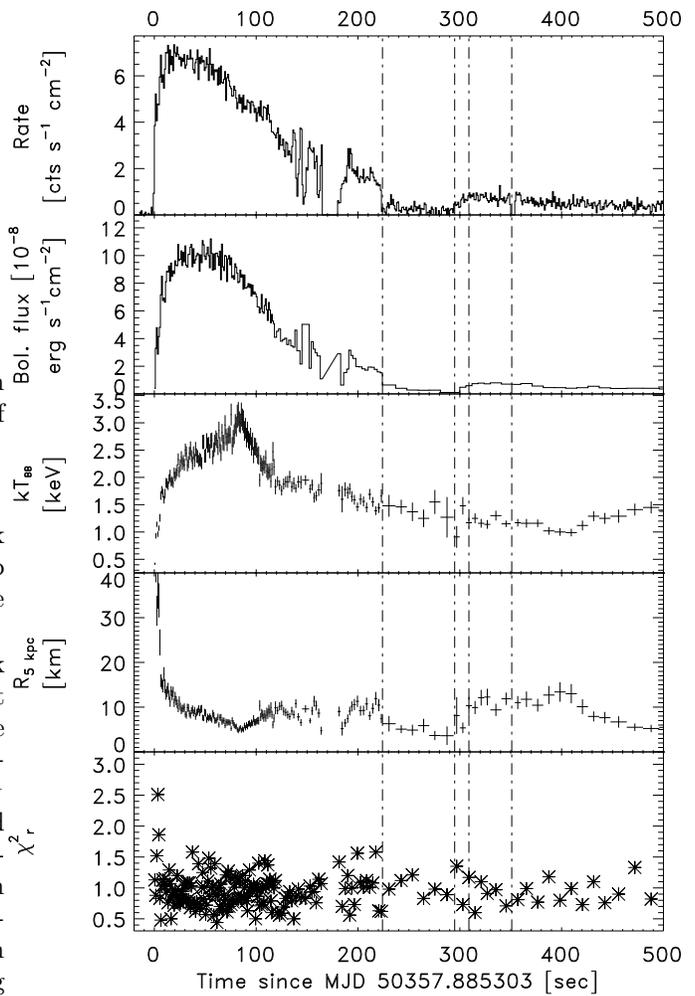}
\caption[]{a. Time history of observed photon flux. b. Bolometric flux
of modeled black body radiation. c. Color temperature of black body
radiation. d. Sphere-equivalent radius of bb radiation for a distance
of 5 kpc (the first data point of $208\pm25$~km is outside the plot
borders). e. $\chi^2_\nu$ of fits. The vertical lines indicate time
intervals for further spectroscopy of the dip (see Fig.~\ref{figdip}).
\label{figburst}}
\end{figure}

\begin{figure}[t]
\includegraphics[width=\columnwidth,angle=0]{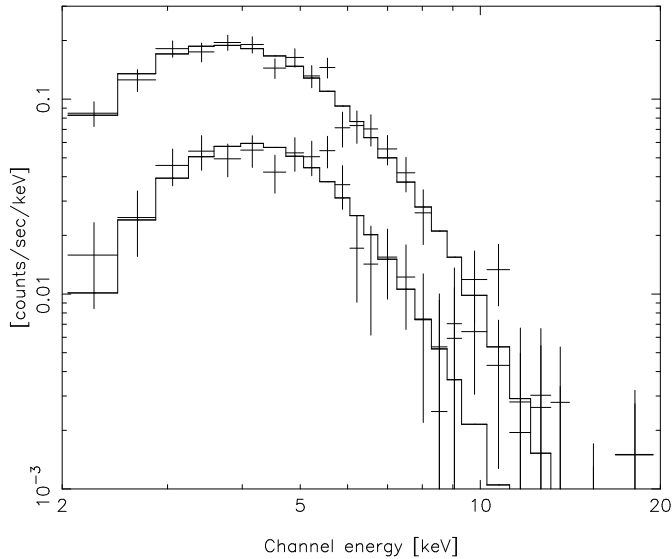}
\caption[]{Spectra of the 71~s lower-flux interval (lower spectrum;
first interval between vertical lines in Fig.~\ref{figburst}) and the
41~s interval after the dip (upper spectrum; last interval in
Fig.~\ref{figburst}). The crosses indicate the measurements (vertical
lengths indicate 1$\sigma$ error intervals; histograms indicate the
fitted model with free $N_{\rm H}$ and fixed radius).
\label{figdip}}
\end{figure}

We modeled the spectrum of the long burst with black body radiation
and present the results in Fig.~\ref{figburst}. The burst starts with
a strong photospheric radius expansion phase which ends after a few
seconds. Subsequently the bolometric flux remains at a level of about
10$^{-7}$~\ecs\ for one minute during which the temperature rises and
the radius shows a slow instead of fast decrease. This suggests that
the photosphere is continuously injected with fresh (radiation)
energy since the fallback time of the atmosphere is much
shorter. Thereafter the burst decays in an ordinary fashion for about
3 minutes with decreasing temperatures and constant radii. This
suggests cooling by a fairly thick layer.  Four minutes after the
burst onset the flux suddenly (within 2 s) drops by a factor of 4. The
flux remains low for 73~s after which it slowly rises during 20~s and
resumes the pre-drop decay. The e-folding decay time then is
$236\pm11$~s. The cause of the dip can be modeled in two ways: either
through a variation of $N_{\rm H}$ ($\chi^2_\nu=0.85$ for $\nu=78$ for
simultaneously fitting the 3 spectra of the dip [exposure time 71 s],
the rise out of the dip [14 s], and a period afterwards [41 s]) or
through a variation of the emission area ($\chi^2_\nu=0.82$ for
$\nu=78$) . Fig.~\ref{figburst} shows the results of the latter model.
The implied reduction in emission area is a factor of 6 (or 2.5 in
radius).  When modeled through an increase of the absorption, the
implied maximum column density is $N_{\rm
H}=8.3^{+2.9}_{-2.5}\times10^{22}$~cm$^{-2}$ (90\% confidence; compare
with $N_{\rm H}=4.2\times10^{21}$~cm$^{-2}$ outside the dip).

If one equalizes the peak bolometric flux to the Eddington limit of a
canonical $1.4~M_\odot$ neutron star, the implied distance is
4.1~kpc for a hydrogen-rich photosphere and 5.4~kpc if it is
hydrogen-poor.

Ignoring the behavior of the source during the dip and data drop outs,
the total bolometric energy output for a distance of 5.4 kpc is
estimated to be $(0.9\pm0.4)\times10^{41}$~erg. This is a factor of
of at least 3 smaller than any of the 13 superbursts observed sofar
(cf, Kuulkers 2004 and In 't Zand et al. 2004a) but similar to the most
energetic non-super X-ray burst (from SLX 1737-282, In 't Zand et
al. 2002).

What is the cause of the 1.2~min drop during the decay phase?  The
profile of the flux history looks strikingly similar to a partial
eclipse. However, this is inconceivable because 1) the radiating
surface is so small that a {\em partial} eclipse seems very unlikely,
and 2) eclipses have never been seen in \bron\ down to very good
limits (e.g., Juett \& Chakrabarty 2003).

The effect looks a bit like the few-second long dip seen in the
superburst from 4U~1820-303 (Strohmayer \& Brown 2002). There it was
attributed to the combination of photospheric radius expansion and a
clean sweep of the normally X-radiating inner accretion disk. This
explanation is inconsistent with the temperature evolution seen in
\bron. Still, it seems likely that the dip is related to a perhaps
more moderate change in the accretion flow geometry induced by the
radiation pressure of the luminous flash. An indication of that is
provided by the strong variability in the minute before the dip.

\section{Flux history of persistent emission}
\label{flux}

\begin{figure}[!t]
\includegraphics[width=\columnwidth,angle=0]{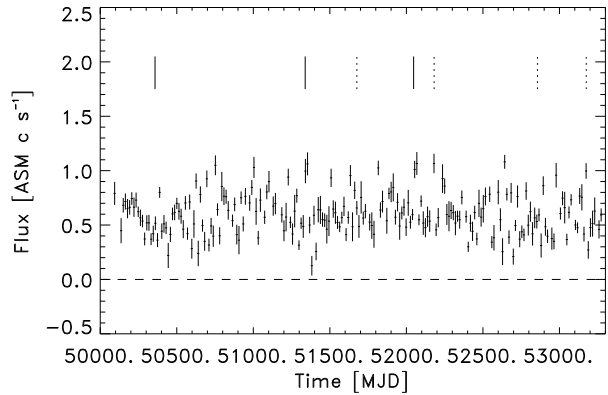}
\caption[]{RXTE/ASM 2-12 keV light curve with bin time of 14~d and
eliminating all data points from SSC3 and with an error in excess of
0.15 c~s$^{-1}$. The vertical lines in the top indicate the times when
bursts were detected (solid lines WFC, dashed lines RXTE PCA (1st and last)
and ASM).
\label{figasm}}
\end{figure}

Given the wide variety of burst durations, it is of interest to test
whether there is a connection with a varying mass accretion
rate. Therefore, we studied data relating to the persistent flux
history. \bron\ is quite faint for monitoring devices such as the WFC
or the RXTE All-Sky Monitor (Levine et al. 1996). This hampers
accurate measurements on time scales below a few weeks.  In
Fig.~\ref{figasm}, the 2-12 keV flux history is plotted as measured
with the RXTE ASM with a 14~d binning time. On that time scale the
flux ranges between 0.3 and 1.0~ASM~c~s$^{-1}$ which, for a Crab-like
spectrum, translates to a 2-10 keV flux of
$(1-3)\times10^{-10}$~\ecs. We note that no flares were observed on
shorter time scales above a limit of roughly 0.1 Crab units, except for
the bursts.  In the same plot the times of the 7 bursts are
indicated. The long (first) burst distinguishes itself from the other
bursts by occurring during a somewhat tranquil low state of the
source.

There has been one broad-band X-ray measurement of the source, by the
BeppoSAX Narrow Field Instruments (Boella et al. 1997a) on 1998 April
22. Broad-band coverage was obtained through the Low-Energy
Concentrator Spectrometer (LECS; Parmar et al. 1997; 0.1-3.0 keV; 16
ksec exposure time), the Medium Energy Concentrator Spectrometer (MECS;
Boella et al. 1997b; 1.6--10.0 keV; 28~ksec) and the Phoswhich Detector
System (PDS; Frontera et al. 1997; 15-200 keV; 12 ksec). The LECS and
MECS provided imaging data, while the PDS operated with a collimator
that rocked between on-source and background pointings 240\arcmin\
from the source position. We verified that no bright X-ray source was
contained in the background pointings.

\begin{figure}[!t]
\includegraphics[height=\columnwidth,angle=270]{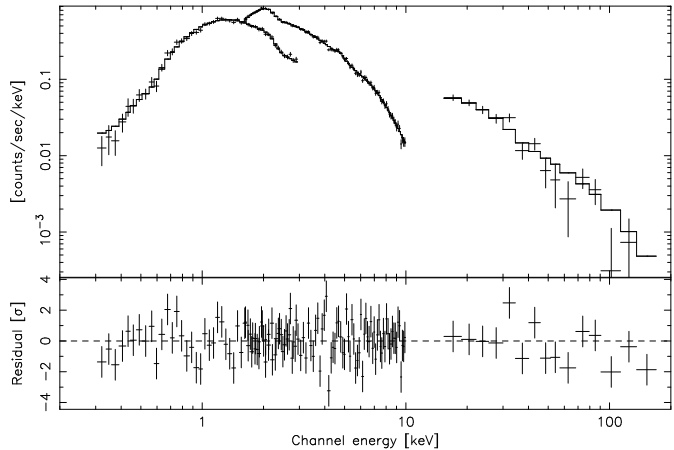}
\caption[]{Top panel: from left to right LECS, MECS and PDS spectrum 
(crosses with vertical lines depicting 1$\sigma$ error bars)
and best fit result for power law and black body model (histogram).
Bottom panel: fit residuals.
\label{nfispectrum}}
\end{figure}

We employed standard extraction and data analysis techniques (e.g., In
't Zand et al. 1999) and restricted further analysis to those photon
energies where there is a significant detection (extending from 0.3 to
120 keV). The LECS and MECS extraction radii were 4\arcmin, a
LECS/MECS and PDS/MECS normalization factor was left free during
spectral fits, LECS and MECS background spectra were determined from
independent long observations on empty fields, and no systematic
uncertainty was included. Various models were tested against the data;
two have a satisfactory result. These are presented in
Table~\ref{tabnfi}. The power-law fit is shown in
Fig.~\ref{nfispectrum}. We tested a pure Comptonized model (model {\tt
comptt} in {\tt XSPEC}; Arnaud 1996; Titarchuk 1994; Hua \& Titarchuk
1995; Titarchuk \& Lyubarskij 1995), and a simple power law. Both
models were absorbed (following the model by Morrison \& McCammon
1983) and a black body component was included describing the 0.7~keV
feature discussed by Juett et al. (2001). The fit results are
consistent with those obtained by Juett \& Chakrabarty (2003). The
(absorbed) 2--10 keV flux is consistent with the ASM measurements.
The unabsorbed 0.1--200 keV flux is the same in both cases.

The 0.1--200 keV flux is of $(6.0\pm0.5)\times10^{-10}$~\ecs\ is less
than 1\% of the bolometric burst peak flux of
$(1.0\pm0.1)\times10^{-7}$~\ecs\ which is thought to be the Eddington
limit.  For a distance of 4.1--5.4~kpc the 0.1--200~keV luminosity is
$(1.2-2.1)\times10^{36}$~\lum.  The ASM light curve suggests that the
source never becomes brighter than roughly twice this value, on time
scales of weeks. The 2-10 keV absorbed flux is also consistent with
similar measurements since the 1970s as compiled by Juett et
al. (2003) which range between 0.9 and 2.7$\times10^{-10}$~\ecs,
except for an ASCA measurement on 1995 May 2 (MJD 49839; 17 months
before the first burst) when the flux was 7.0$\times10^{-10}$~\ecs.
We note that an analysis of the near-to-continuous BATSE data set on
\bron\ by Harmon et al. (2004) does not discuss a peak in 1995.

\begin{table}[!t]
\caption[]{Spectral parameters of acceptable model fits to the NFI
spectrum. $\Gamma$ is the photon index. Errors are for 90\% 
confidence.\label{tabnfi}}
\begin{tabular}{ll}
\hline
Model                    & power law + black body\\
$N_{\rm H}$              & $(3.1\pm0.3)\times10^{21}$\\
bb $kT$                  & $0.51\pm0.03$ keV \\
$\Gamma$                 & $2.07\pm0.05$ \\
$\chi^2_\nu$             & 1.21 (133 dof) \\
Unabs. 0.1--200 keV flux & $(6.4\pm0.5)\times10^{-10}$~\ecs \\
Abs. 2--10 keV flux      & $(1.37\pm0.02)\times10^{-10}$~\ecs \\
\hline
Model                    & comptt + black body \\
$N_{\rm H}$              & $(2.8\pm0.3)\times10^{21}$ \\
bb $kT$                  & $0.51\pm0.03$ keV \\
$kT_{\rm plasma}$        & $34_{-17}^{+110}$~keV \\
$\tau$ (spherical geometry) & $1\pm0.5$ \\
$\chi^2_{\rm r}$         & 1.20 (131 dof) \\
Unabs. 0.1--200 keV flux & $(5.5\pm0.5)\times10^{-10}$~\ecs \\
Abs. 2--10 keV flux      & $(1.37\pm0.02)\times10^{-10}$~\ecs \\
\hline
\end{tabular}
\end{table}

\bron\ has never been seen in an off state, despite extensive
coverage since the early 1970s. It is therefore not an X-ray
transient. The reason that it is persistent while the luminosity is
rather low is possibly related to the presumed ultracompact nature. A
smaller orbit generally implies a smaller accretion disk. Therefore,
the disk will remain completely photo-ionized at lower accretion rates
and the accretion will sustain all the way to the neutron star rather
than turn off due to an accretion disk instability (White, Kaluzienski
\& Swank 1984; van Paradijs 1996; Deloye \& Bildsten 2003).

\section{Discussion}
\label{discussion}

\subsection{Short and long helium bursts}
\label{dis1}

In general, the longevity of an X-ray burst is determined by the
duration of the nuclear burning, and by the thickness and composition
(through thermal conductivity) of the layer where the burning deposits
heat. Whereas helium and carbon burn very rapidly, hydrogen burning
involves slow beta decays, and so can prolong the energy
generation. Slow hydrogen burning via the rp-process (Wallace \&
Woosley 1981) is believed to power the minutes long tails of bursts
from GS~1826-24 (Galloway et al. 2004). For accretion rates $\sim 0.1$
times Eddington, appropriate for most X-ray burst sources, this has
led to the identification of ``short'' duration bursts ($\sim 10$~s)
with helium-dominated flashes, and long duration ($\sim 100$~s) bursts
with hydrogen-dominated flashes. In this picture, the long duration
burst from \bron\ is difficult to explain because we expect the
companion to be hydrogen deficient. However, at low accretion rates,
long bursts can arise because of very thick fuel layers that
accumulate between bursts. These thick layers have a long cooling
time, leading to long burst durations. In this section, we show that
the long burst from \bron\ is naturally explained by accretion of pure
helium at the observed rate of $0.01$ of the Eddington accretion
rate.

\begin{figure}[!t]
\includegraphics[width=\columnwidth,angle=0]{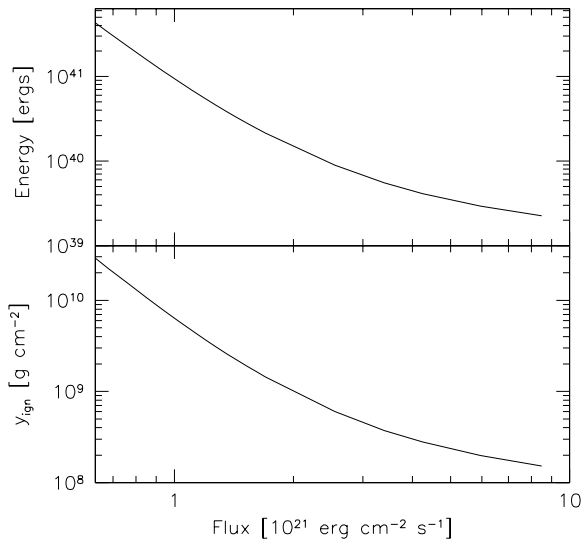}
\caption[]{Ignition thicknesses and flash energies as a function of
heat flux.
\label{figign}}
\end{figure}

The observed burst energy of $E_{\rm nuc}=10^{41}\ {\rm ergs}$
implies an ignition column depth of $y=E_{\rm nuc}(1+z)/4\pi R^2Q_
{\rm nuc}\approx 7\times 10^9\ {\rm g\ cm^{-2}}$, where $Q_{\rm nuc}
\approx 1.6$ MeV per nucleon is the energy release for helium burning
to iron group nuclei, $R$ is the neutron star radius, and $z$ is the
gravitational redshift (we assume $R=10\ {\rm km}$ and $z=0.31$,
appropriate for a $1.4~M_\odot$ neutron star). We have calculated the
ignition depth for pure helium following the ignition calculations of
Cumming \& Bildsten (2000). The calculation involves finding the
temperature profile of the accumulating fuel layer, and adjusting the
layer thickness until the criterion for unstable ignition is met at
the base. Since hydrogen burning is not active for pure helium
accretion, the temperature profile of the layer is set by the heat
flux emerging from the neutron star crust. We write this heat flux as
$\dot mQ_b\ {\rm erg\ cm^{-2}\ s^{-1}}$, where $\dot m$ is the mass
accretion rate per unit area, and $Q_b$ is the energy per gram
released in the crust by pycnonuclear reactions that flows outwards.
For low $\dot m$, Brown (2000) found that almost all of the $\approx
1.4$ MeV per nucleon released in the crust comes out through the
surface\footnote{We expect that the value of $Q_b$ will depend on the
thermal properties of the neutron star interior, for example, the core
temperature and crust thermal conductivity. We will investigate the
dependence of the ignition conditions on these factors in a future
paper.} (see Fig.~11 of Brown 2000). For $Q_b\approx 1$ MeV per
nucleon or $Q_b\approx 10^{18}\ {\rm erg\ g^{-1}}$, and using the
Eddington accretion rate $\dot m_{\rm Edd}\approx 10^5\ {\rm g\ cm^
{-2}\ s^{-1}}$, we find $F_b\approx 10^{21}\ {\rm erg\ cm^{-2}\ s^
{-1}}$ for accretion at $\dot m=0.01\ \dot m_{\rm
Edd}$. Figure~\ref{figign} shows the ignition column depth and
predicted burst energy as a function of base flux. For a base flux of
$10^{21}\ {\rm erg\ cm^{-2}\ s^{-1}}$, we find an ignition column
depth of $y\approx 10^{10}\ {\rm g\ cm^{-2}}$, in good agreement with
the value inferred from the burst energetics.

\begin{figure}[!t]
\includegraphics[width=\columnwidth,angle=0]{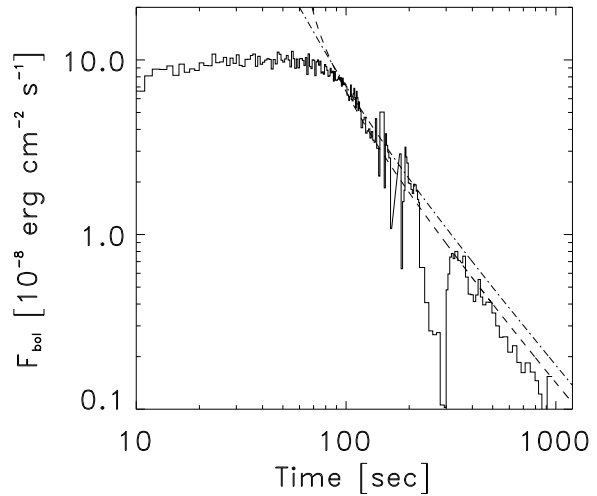}
\caption[]{Comparison of the observed decay of the bolometric black
body flux (histogram) with a theoretical model (Cumming \& Macbeth
2004) for the cooling rate of a column of depth
$7\times10^{9}$~g~cm$^{-2}$ and a nuclear energy release of
$1.6\times10^{18}$~erg~g$^{-1}$ which is expected for helium burning
to iron (dashed curve). To illustrate the dependence on these two
parameters a model is shown with the same energy output
(10$^{41}$~erg) but for a column depth of $1\times10^{10}$~g~cm$^{-2}$
and an energy release of $1.0\times10^{18}$~erg~g$^{-1}$
(dashed-dotted curve).
\label{figprof}}
\end{figure}

Additional constraints come from the burst lightcurve and the
recurrence time. Figure~\ref{figprof} shows the observed lightcurve
compared with theoretical cooling models calculated following Cumming
\& Macbeth (2004). In these models, the burning is assumed to take
place instantaneously, since helium burning is extremely rapid. We
then follow the cooling of the hot layer using a time-dependent
thermal diffusion code. We show two curves with a total energy release
of $10^ {41}\ {\rm ergs}$, with column depths $7\times 10^9$ and
$10^{10}\ {\rm g\ cm^{-2}}$. The observed decay is well-reproduced by
these models. Unfortunately, the recurrence time of the long burst is
not well constrained by observations although the suggestion is that
it is long. The expected recurrence time from the ignition models is
$y/\dot m=116\ {\rm days}\ (y/10^{10}\ {\rm g\ cm^{-2}})(\dot m/10^3\
{\rm g\ cm^{-2}\ s^{-1}})^{-1}$. \bron\ was almost continuously
observed with the WFCs for 4 days prior to the burst, but there were
two data gaps so that the lower limit to the recurrence time is only
1.1 days. In the 87.3~d period prior to the long burst, 15.6~d of
effective exposure time were collected with no burst detections. The
only other X-ray experiment with coverage of \bron\ in 1996 is the
All-Sky Monitor on RXTE, also without burst detections.

We have assumed that the accreted material is pure helium.  If
the accreted material contains carbon, carbon ignition is also
possible.  However, the burst energy in that case would be $\gg
10^{42}\ {\rm ergs}$, and recurrence time $>10$ years (Cumming \&
Bildsten 2001).  If hydrogen is included in the accreted material,
additional heating arises because of CNO burning. Our steady-state
accumulation models are only applicable if the temperature is large
enough during accumulation ($\ga 8\times 10^7\ {\rm K}$) that the
temperature-independent hot CNO cycle operates. Assuming this to be
the case, we find that accretion of solar composition material at a
rate $10^3\ {\rm g\ cm^{-2}\ s^{-1}}$ gives an ignition depth smaller
than the pure helium case, $6\times 10^8\ {\rm g\ cm^{-2}}$, and
energy release $\approx 10^{40}\ {\rm ergs}$ (in this model, only the
upper 10\% by mass of the layer contains hydrogen). In fact, depending
on the accretion history, CNO burning may be much less than the hot
CNO value. Narayan \& Heyl (2003) calculate steady-state models with
detailed CNO burning, and find ignition column depths of $\approx 3
\times 10^9\ {\rm g\ cm^{-2}}$ for $\dot m=0.01\ \dot m_{\rm Edd}$.
Therefore, the burst properties may be similar to those observed if
some hydrogen is present. However, we do not expect this because of
the likely ultracompact nature of this source.

If the long burst is indeed a helium burst and the other bursts are as
well, then the widely varying burst duration must be directly related
to the layer thickness. The ignition condition predicts that the
ignition thickness decreases with increasing temperature. This implies
that the temperature increases from burst 1 to 4 and then decreases
again. If the hot CNO cycle is not active, the heat flux from the
core/crust and the composition of the outer 100~m of the NS determines
the temperature (Brown, Bildsten \& Chang 2002). Since the time scale
of variation of the crust temperature is expected to be much longer
than the burst interval time (years rather than months), the
suggestion is there that the composition of the layer changes between
bursts. Given the limited accuracy of our measurements we are unable
to test this quantitatively. Temperature variability of roughly a
factor of 2 to 3 would be needed to explain a variety of burst
durations of a factor of 10 to 100. The ASM data suggest that the
accretion rate shows more variability during the six shorter
bursts. Perhaps this explains the earlier ignition and presence of
shorter (less energetic) bursts, much as was observed in KS 1731-260
by Cornelisse (et al. 2003; 2004).

\subsection{Evolutionary considerations}
\label{dis2}

Since the surface layers of the donor are the source of the matter
flowing through the accretion disk onto the neutron star, we may
conclude that these surface layers are deficient in hydrogen (from the
optical spectrum of the disk and the presumed ultracompact
nature), do contain helium (from the X-ray bursts), and have an
enhanced Ne/O abundance ratio (from the X-ray spectrum). This
information is useful to discriminate between different evolutionary
scenarios.

The evolutionary path of a binary in which a neutron star accretes
matter from a companion depends to a large extent on the evolutionary
state of the donor at the moment at which mass transfer starts. If
mass transfer starts on the main sequence, the orbit shrinks to a
minimum of around 70-80 min, and then expands again. At all times
during the evolution, the transferred mass consists mainly of
hydrogen. It has been suggested that strong magnetic braking may cause
the orbit to shrink even if mass transfer is initiated after the donor
has evolved a little beyond the terminal age main sequence (Tutukov et
al.\ 1985). The decrease of the orbital period may then proceed to
periods less than 70-80 min before the orbit expands again. Even
though the hydrogen contents of the transferred mass drop at the
shortest periods, the transferred mass is hydrogen rich throughout the
evolution in this case also (see e.g.\ Tables 1, 2 in Van der Sluys et
al.\ 2005a). Van der Sluys et al.\ (2005a,b) show that this path to
ultrashort periods demands both very special initial conditions and
very strong magnetic braking, and thus is unlikely to be important.

If mass transfer starts during shell burning as the star ascends the
giant branch, it will lead to long periods if the mass transfer is
stable, and eventually to a wide binary of a neutron star an an
undermassive white dwarf (e.g.\ Webbink et al.\ 1983). However, if the
mass transfer is unstable, a spiral-in may ensue, and lead to a close
binary of the core of the giant and the neutron star. Mass transfer is
increasingly likely to be unstable if the donor star has higher mass,
and is further evolved along the giant branch and hence its
envelope is fully convective. The post-spiral-in close binary evolves
to even shorter periods through loss of angular momentum via
gravitational radiation, which may bring the core of the giant, by
then cooled into a white dwarf, into contact with the Roche lobe,
after which mass transfer starts again.  If the spiral-in started with
the donor in a phase of hydrogen shell burning, the white dwarf is a
helium white dwarf. Donors in a phase of helium and carbon shell
burning would lead to carbon-oxygen and neon-magnesium-oxygen white
dwarfs, respectively.  Mass transfer from the white dwarf to the
neutron star is dynamically unstable if the white dwarf has a mass
which is too high.  The precise limit is somewhat uncertain,
depending on the amount of mass and angular momentum loss, but is
probably near 0.4-0.5 $M_\odot$ (see e.g.\ Yungelson et al.\
2002). This limit excludes neon-magnesium-oxygen white dwarfs as well
as the more massive carbon-oxygen white dwarfs.  Only low-mass
carbon-oxygen white dwarfs and helium white dwarfs are possible stable
donors for a neutron star.

Mass transfer in a system with a white dwarf donor is a very strong
function of the mass of the white dwarf. Immediately after contact, at
a period on the order of a few minutes, the mass transfer is highly
super-Eddington, and the white dwarf mass decreases rapidly. The orbit
expands, and the donor mass decreases quickly until the binary has a
mass transfer rate that is sufficiently low to be sustained for a
longer period of time. The shortest orbital period observed for a
system with a neutron star and white dwarf donor is 11\,min. A rough
estimate of the mass of the donor $M$ as a function of orbital period
$P_b$ can be made by combining the mass-radius relation of a white
dwarf with the equation giving the size of the Roche lobe for the less
massive star in a binary: this gives $M/M_\odot \sim (50~{\mathrm
s}/P_b)$. Thus, the 11\,min binary has a donor with mass less than
$0.1M_\odot$ (Verbunt 1987), and the other ultrashort period systems
with known orbital periods have donors of smaller masses still. This
implies that the composition of the mass being transferred to the
neutron star in these binaries is that of the innermost material of
the initial white dwarf.

\begin{figure}[!t]
\includegraphics[width=\columnwidth,angle=0]{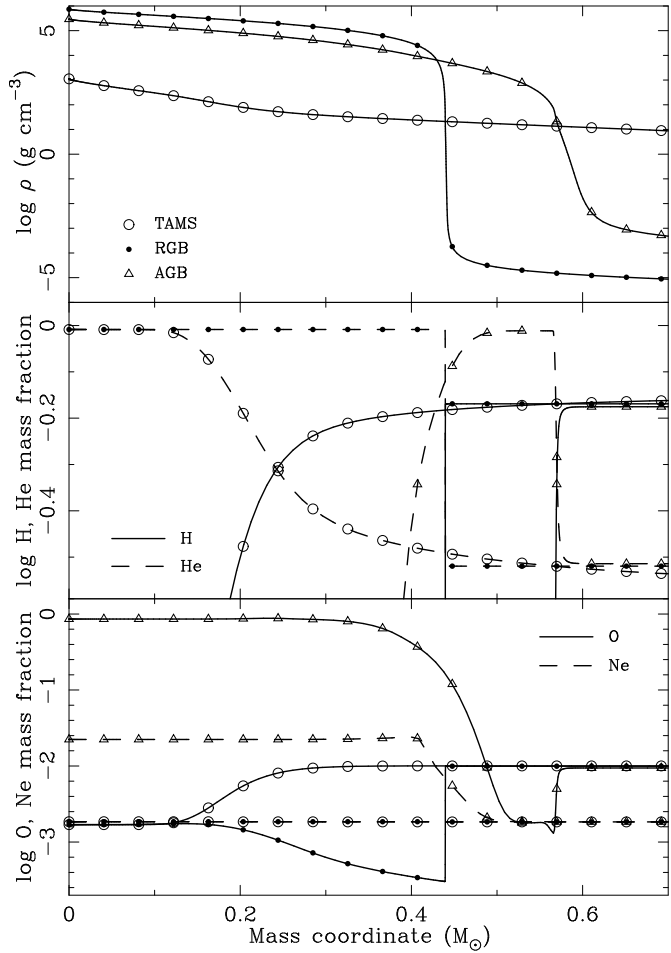}
\caption[]{ Three structure plots for different moments in the
 evolution of a 1.5\,$M_\odot$ model.  Only the central 0.7\,$M_\odot$
 is shown, as a function of the mass coordinate.  {\it Upper panel}
 {\bf a)}: Logarithm of the mass density.  {\it Middle panel} {\bf
 b)}: Logarithm of the hydrogen (solid lines) and helium (dashed
 lines) mass fraction .  {\it Lower panel} {\bf c)}: Logarithm of the
 oxygen (solid lines) and neon (dashed lines) mass fraction.  The
 symbols on the lines indicate the model, as shown in the upper panel.
\label{fig:structure}
}
\end{figure}

We illustrate the compositions of the helium and carbon-oxygen white
dwarfs by means of the core of a model star of 1.5~$M_\odot$ in
Fig.~\ref{fig:structure} and Table\,\ref{tab:structure}.  These
compositions were computed with the evolution code of Eggleton (1971,
1972) with updated input physics as described in Pols et al.\ (1995).
We do not expect mass loss to have any effect, because the profiles in
the inner core have already been established by the time mass loss
becomes important. Mass transfer from the white dwarf donor can cause
its mantle to become convective, especially near and beyond the period
minimum.  However, since the core of the star is very homogeneous
(Fig.\,\ref{fig:structure}), this has little influence on the surface
abundances. The demand that the mass transferred in 2S\,9018$-$549
contains helium is obviously compatible with a donor consisting of the
inner $\ltap0.1M_\odot$ of an initial helium white dwarf; however, it
is not compatible with a donor consisting of the inner
$\ltap0.1M_\odot$ of an initial carbon-oxygen white dwarf. As shown by
Fig.\,\ref{fig:structure} the helium content of the central mass of a
carbon-oxygen white dwarf is zero (it is at the minimum allowed for
computational stability in the evolution code, $10^{-12}$). From this,
one would have to conclude that the donor at the onset of mass
transfer was a helium white dwarf, rather than a carbon-oxygen white
dwarf.

This conclusion is strengthened when we consider the neon and oxygen
abundances. In the helium core, the neon abundance is still at the
zero-age main-sequence composition of the progenitor star, not
affected by nuclear evolution (Fig.\,\ref{fig:structure}).  The oxygen
abundance, however, is lower in the helium core as oxygen is converted
into nitrogen in the CNO cycle (Iben 1967). The depletion of
oxygen is stronger in more massive progenitors, because the CNO cycle
takes place at higher temperatures (see Table~\ref{tab:masses}). Thus
the Ne/O abundance ratio is predicted to be high if the donor in
2S\,9018$-$549 consists of the central mass of a helium white
dwarf. We point out that there is no observational evidence that the
mass fraction of neon is enhanced since a complete measurement of the
abundance of all expected elements, particularly helium and carbon, is
lacking; only the abundance ratios Ne/O and Ne/Fe have been measured
(Juett et al. 2003). Neither is there conclusive observational
evidence for an increased oxygen abundance, even when considering the
optical spectrum which does suggest the presence of oxygen (and
carbon) lines but with insufficient significance to prove the presence
(Nelemans et al.  2004).

In a carbon-oxygen white dwarf both neon and oxygen are more abundant,
but oxygen more so than neon (Fig.\,\ref{fig:structure}). The increase
in neon abundance is caused by the conversion of $^{14}$N, produced by
the CNO cycle, into $^{22}$Ne during core He-burning.\footnote{The
nuclear network in the evolution code does not follow the $^{22}$Ne
abundance directly, but instead the burnt $^{14}$N is added to the
$^{20}$Ne abundance assuming particle number conservation (see Pols et
al.\ 1995). The total neon mass fraction we find is therefore slightly
underestimated.}  However, this is dwarfed by the production of oxygen
by helium burning.  The possibility exists, if the white dwarf has
time to cool enough for crystallization to take place, that $^{22}$Ne
settles in the centre (Yungelson et al. 2002) and reaches there the
so-called azeotropic mass fraction, which is between 0.05 and 0.09
(Isern et al. 1991). This is not enough, however, to increase the
Ne/O ratio above the solar value. Taking into account that the
azeotropic Ne abundance may be underestimated by up to a factor three,
the Ne/O ratio might be barely reconciled with the observed value, as
noted by Yungelson et al. (2002).  Nevertheless, if the donor in
\bron\ consists of the central mass of a carbon-oxygen white dwarf the
Ne/O abundance ratio is expected to be rather low. Conversely, the
Ne/O ratio observed for \bron\ is naturally explained if its donor is
the central mass of a helium white dwarf, rather than a carbon-oxygen
white dwarf.

\begin{table}
\caption{ Properties of the 1.5\,$M_\odot$ model at the moments the
structure plots of Fig.\,\ref{fig:structure} were made.  The age is in
Gyr, the helium and carbon-oxygen core masses in $M_\odot$.  The last
two columns give the mass fraction ratio Ne/O in the core, and the
ratio of this number to the initial (ZAMS or ISM) Ne/O ratio.  }
\label{tab:structure}
\begin{tabular}{ccccccc}
 \hline
 \hline
  Model & Age & M$_{\rm He}$  & M$_{\rm CO}$  & (Ne/O)$_{\rm c}$ & (Ne/O)$_{\rm c}$/(Ne/O)$_{\rm in}$ \\
 \hline
 TAMS & 2.5949 & 0.166 & 0.000 & 1.093 &  5.97 \\
 RGB  & 2.8495 & 0.440 & 0.000 & 1.093 &  5.97 \\
 AGB  & 2.9852 & 0.568 & 0.379 & 0.0260 & 0.142 \\
 \hline
\end{tabular}
\end{table}

The helium core denuded by a spiral-in undergoes helium burning if its
mass is higher than about 0.34\,$M_\odot$ before it becomes
degenerate, which will be the case for stars with initial mass higher
than 2.25\,$M_\odot$ (see Table\,\ref{tab:masses}).  In this case a
hybrid white dwarf may be formed, with a carbon-oxygen core and a
helium mantle. When this white dwarf transfers mass to a neutron star,
it will rapidly lose its helium mantle; at orbital periods in excess
of 11\,min no helium is left.

\begin{table}[t]
\caption{ Helium core masses and core abundances for model stars with
masses between 1.0 and 5.0 $M_\odot$.  The second column gives the
range of helium core masses (in $M_\odot$) that are obtained between
the formation of the core and core helium ignition.  The third and
fourth column give the Ne/O abundance of the core, relative to the
initial (ISM) Ne/O abundance, for the helium core (RGB) and
carbon-oxygen core (AGB) respectively.  }
\label{tab:masses}
\begin{center}
\begin{tabular}{cccc}
 \hline
 \hline
  M & M$_{\rm He-core}$ & \multicolumn{2}{c}{(Ne/O)$_{\rm c}$/(Ne/O)$_{\rm in}$} \\
    &                       &                     RGB & AGB \\
 \hline
 1.00 & 0.00 -- 0.47 & 2.25 & 0.14 \\
 1.50 & 0.15 -- 0.47 & 5.97 & 0.15 \\
 2.00 & 0.23 -- 0.39 & 9.03 & 0.14 \\
 2.25 & 0.30 -- 0.34 & 10.2 & 0.13 \\
 2.50 & 0.34 -- 0.36 & 11.3 & 0.13 \\
 3.00 & 0.38 -- 0.42 & 12.9 & 0.14 \\
 4.00 & 0.63 -- 0.64 & 16.3 & 0.15 \\
 5.00 & 0.85 -- 0.85 & 19.0 & 0.16 \\
 \hline
\end{tabular}
\end{center}
\end{table}

Thus, a helium white dwarf is the most promising donor in the
2S\,9018$-$549 system. Such a donor is formed when a star with initial
mass less than 2.25\,$M_\odot$ enters a spiral-in phase on its first
ascent of the giant branch. Based on the work by Deloye \& Bildsten
(2003), the observed mass accretion rate predicts for a He WD donor an
orbital period between 25 and 30 minutes. These results are in general
agreement with calculations by Belczynski \& Taam (2004) who predict
that of all ultracompact binaries with a neutron star accretor, 60\%
may have a helium white dwarf donor. Furthermore, our calculations
suggest that AM CVn systems, ultracompact binaries in which the
accretor is a white dwarf, should show an enhanced Ne/O ratio if they
have a helium white dwarf companion. Indeed, such an enhancement was
recently observed in the AM CVn system GP Com (Strohmayer 2004).

\section{Conclusion}

In conclusion, we have shown that the properties of the long
X-ray burst from \bron\ and the enhanced Ne/O abundance ratio are both
consistent with the companion star being a helium white dwarf.  This
scenario would seem to be at odds with the presence of C and O lines
and the absence of He lines in the optical spectrum. However, the
evidence for C and O lines is inconclusive. Nelemans et al.~(2004)
find that the case is less clear for \bron\ than for 4U~0614+091 and
4U~1543-624 due to the relative faintness of the optical counterpart,
and confirmation through deeper observations would be desirable. In
addition, non-LTE effects may be important. This is the case in the
UV, where Werner et al.~(2004) calculated accretion disk models for
the UV spectrum and found the He II 1640A line depth to be rather weak
even for large helium abundances. Therefore, as yet we regard all
X-ray and optical measurements of \bron\ to be consistent with a
helium white dwarf donor star.

\acknowledgement 

We thank Ron Remillard for providing the high time-resolution data of
the two bursts detected with the RXTE/ASM, and Lars Bildsten, Deepto
Chakrabarty, Edward Brown, Duncan Galloway, Peter Jonker and Gijs
Nelemans and for useful discussions. JZ acknowledges support from the
Netherlands Organization for Scientific Research (NWO). AC
acknowledges support from McGill University startup funds, an NSERC
Discovery Grant, Le Fonds Qu\'eb\'ecois de la Recherche sur la Nature
et les Technologies, and the Canadian Institute for Advanced
Research. Gerrit Wiersma, Jaap Schuurmans, Nuovo Telespazio and the
ASI Science Data Center are thanked for continued support.

\end{document}